\documentclass[12pt,letterpaper]{JHEP3}
\usepackage{amsmath,epsf,amssymb,latexsym,cite,graphics,bbm}
\usepackage[matrix,arrow,frame,import,curve,color]{xy}

\DeclareFontFamily{U}{rsf}{}
\DeclareFontShape{U}{rsf}{m}{n}{
  <5> <6> rsfs5 <7> <8> <9> rsfs7 <10-> rsfs10}{}
\DeclareMathAlphabet\Scr{U}{rsf}{m}{n}

\def\p{\partial}

\def\la{\langle}
\def\ra{\rangle}

\def\ff#1#2{{\textstyle\frac{#1}{#2}}}

\def\cA{{\cal A}}

\def\cE{{\cal E}}

\def\cL{{\cal L}}

\def\cW{{\cal W}}

\def\ep{{\epsilon}}

\newcommand\gh{\widehat{g}}

\newcommand\gt{\widetilde{g}}

\newcommand\Rt{\widetilde{R}}

\newcommand\nablah{\widehat{\nabla}}
\newcommand\nablat{\widetilde{\nabla}}

\title{A Non-Relativistic Weyl Anomaly}
\author{Ido Adam, Ilarion V.~Melnikov and Stefan Theisen\\
\normalsize Max-Planck-Institut f\"ur Gravitationsphysik (Albert-Einstein-Institut),\\
\normalsize Am M\"uhlenberg 1, D-14476 Golm, Germany
}

\abstract{We examine the Weyl anomaly for a four-dimensional $z=3$
  Lifshitz scalar coupled to Ho{\v r}ava's theory of anisotropic
  gravity.  We find a one-loop break-down of scale-invariance at
  second order in the gravitational background.}

\preprint{AEI-2009-061}

\begin{document}


\section{Introduction}
Recently, Ho{\v r}ava constructed an intriguing theory of gravity in four dimensions~\cite{Horava:2009uw}.  Motivated by models for condensed matter systems that exhibit anisotropic scaling phenomena, this Ho{\v r}ava-Lifshitz gravity (HLG) lacks four-dimensional Lorentz invariance but seemingly has better UV properties than Einstein gravity.  A neat aspect of HLG is that for a particular choice of parameters the theory has a classical Weyl gauge symmetry while remaining second order in the time derivatives.  This leads to a natural question:  is the Weyl invariance anomalous?

In this note we study a technically simpler but related question:  a matter system exhibiting appropriate anisotropic scaling may be coupled to HLG in a Weyl-invariant fashion;  is there a Weyl anomaly for the matter system in the HLG background?  We study the problem for a $z=3$ free Lifshitz scalar in $d=4$ and find an anomaly.  

To describe our result, it is useful to recall the work of~\cite{Deser:1993yx} on the classification of Weyl 
anomalies in Lorentz-invariant theories in even dimensions.  Consider the renormalized effective 
action $\cW[g;\mu]$ obtained by integrating out the matter fields.  Here $g$ is the background metric and $\mu$ is 
a mass scale which may arise in removing logarithmic divergences.  The Weyl anomaly is given by
\begin{equation}
\cA (x) \equiv \left. \frac{\delta}{\delta \varphi(x)} \cW[ e^{2\varphi} g; \mu]\right|_{\varphi=0}.
\end{equation}
Following the terminology of~\cite{Deser:1993yx}, dilatation-invariant (i.e. $\mu$-independent) terms in $\cW$ give rise to ``type A'' terms in $\cA$, while $\mu$-dependent terms give rise to ``type B'' anomaly terms.  Recall that
in two dimensions there is only a type A anomaly, while in four dimensions the result is
\begin{equation}
\label{eq:4danom}
\cA|_{d=4} = -a (\text{Euler}) +  c (\text{Weyl})^2 + b \Box R.
\end{equation}
The first of these is of type A, the second is the type B anomaly, and the third may be removed by a local counter-term. 

These anomalies may be uncovered by studying $n$-point functions of the energy momentum tensor in flat space. The 
$d=2$ case is a text-book example (see, e.g. ~\cite{Green:1987sp}), where the study of the two-point function $\la T_{\mu\nu}(x) T_{\rho\sigma}(0)\ra$ quickly leads to the anomaly.  In four dimensions, since the non-trivial terms on the right-hand side of (\ref{eq:4danom}) are quadratic in curvature, a study of three-point functions is necessary to directly determine $a$ and $c$.  However, the scale dependence of $\cW[g;\mu]$ is apparent at the level of the two-point functions. 
To see this, note that the scale dependent term has a schematic form 
\begin{equation}
\cW[g;\mu] = c \int d^4 x~\sqrt{g} C \log(\Box/\mu^2) C,
\end{equation}
where $C$ is the Weyl tensor and $\Box$ is the covariant Laplace operator.
It follows that
\begin{equation}
\mu \frac{d}{d\mu} \la T_{\mu\nu}(x) T_{\rho\sigma} (0) \ra  = c \Delta_{\mu\nu\rho\sigma} \delta^{(4)}(x),
\end{equation}
where $\Delta_{\mu\nu\rho\sigma}$ is a fourth-order derivative operator consistent with $T^\mu_\mu = 0$, $\p^\mu T_{\mu\nu}=0$ and Bose symmetry.   Indeed, as shown in~\cite{Osborn:1993cr}, a naive position-space computation of the two-point function yields a singular distribution, and the expected scale dependence may be uncovered by carefully regularizing this distribution.  Thus, the coefficient of the B-type anomaly may already be extracted from the two-point function.

We have studied the two-point functions of the conserved quantities for the $z=3$ Lifshitz scalar.  We find that while no type A terms arise at this order in the background, a logarithmic divergence leads to a scale dependence of the two-point functions and thus to a type B anomaly.

Our result, while perhaps not very surprising, merits attention for several reasons.  First, it bears on the UV behavior of HLG theories coupled to matter.  As observed in~\cite{Horava:2009uw}, to achieve the desired UV improvements, it is necessary to work at points of enhanced gauge symmetry.  Classically, the Weyl-invariant point is one such choice, and our results rule it out quantum mechanically once HLG is coupled to a scalar.  Second, our work is an illustration of the complexity associated to even the simplest computations in anisotropic theories of gravity, and we hope that the techniques developed herein may be of use in further investigations.  Finally, the Weyl anomaly in a Lorentz-invariant theory is well-known to contain important physical information; our work may be viewed as a first step in learning how to quantify and extract this information in anisotropic theories.

The rest of the paper is organized as follows.  In section~\ref{s:freescalar}, we introduce the $z=3$ free Lifshitz scalar, compute its propagator and couple it to the Weyl-invariant HLG theory.  In section~\ref{s:regularization}, we develop some position-space techniques and test these in the familiar relativistic $d=2$ and $d=4$ examples.  Finally, in section~\ref{s:anomaly}, we apply these techniques to $2$-point functions of conserved quantities in the $z=3$ theory and find the scale dependence symptomatic of a type B anomaly.  We conclude with a brief discussion of our results and some observations on possible type A anomaly terms that follow from Wess-Zumino consistency conditions.

\section{The Lifshitz scalar in a gravitational background} \label{s:freescalar}
\subsection{The free scalar in flat space}
Our starting point is the free scalar with dynamical exponent $z$ in $d=4$.  The Euclidean action is 
\begin{equation}
S =  \frac{1}{2} \int d\tau d^3 x~ \left\{\dot{\phi}^2  + \phi (-\p^2)^z \phi\right\},
\end{equation}
where $\p^2 = \p_i \p_i$ is the spatial Laplacian, and $\dot{\phi} =\p_\tau \phi$.  The propagator for $\phi$ has a simple Fourier space
representation.  Defining $\Delta(\tau,x)$ as a solution to
\begin{equation}
(-\p_\tau^2+(-\p^2)^z) \Delta (\tau,x) = \delta(\tau)\delta^{(3)}(x) \equiv \delta^{(4)}(\tau,x),
\end{equation}
we see that
\begin{equation}
\Delta(\tau,x)  = \int \frac{d\omega d^3 k}{(2\pi)^4} \frac{e^{i\omega\tau +i k x}}{\omega^2 + (k^2)^z}.
\end{equation}
For our purposes, an explicit position-space representation will be useful.  Performing the $\omega$ and angular integrations
leads to
\begin{equation} \label{eq:HL-scalar-propagator}
\Delta(\tau,x) = \frac{|\tau|^{\frac{z-3}{z}}}{4\pi^2 z} \int_0^\infty dv~ v^{\frac{3-2z}{z}} e^{-v} \frac{\sin (v^{\frac{1}{z}} u^{\frac{1}{2}})}{v^{\frac{1}{z}} u^{\frac{1}{2}}}, 
\end{equation}
where $u = x^2 (\tau^2)^{-1/z}$.
Performing the remaining integral, we have
\begin{equation}
\Delta(\tau,x) =  \frac{|\tau|^{\frac{z-3}{z}}}{4\pi^2 z} \sum_{n=0}^{\infty} \frac{\Gamma(\frac{3+2n}{z}-1)}{\Gamma(2n+2)} (-u)^n.
\end{equation}
Plugging in $z=1$, we find the usual propagator for a relativistic massless particle.  Our interest is in the more exotic limit of $z=3$, which leads to
\begin{equation}
\label{eq:prop}
\Delta(\tau,x)|_{z=3} = \frac{1}{12\pi^2} \left[ G(u) -\frac{1}{2} \log(\tau^2 m^2) \right],
\end{equation}
where $m$ is a scale introduced to make sense of the logarithm and to absorb an infinite constant; 
$u = x^2 (\tau^2)^{-1/3}$, and 
\begin{equation}
G(u) = \sum_{n=1}^\infty \frac{\Gamma(\frac{2n}{3})}{\Gamma(2n+2)} (-u)^n.
\end{equation}
$G(u)$ is an analytic function in the complex plane.  As we will see, the relative complexity of $\Delta(\tau,x)$ at $z=3$ makes even free field theory computations a bit more laborious than in the familiar $z=1$ case.  We also note that $z=3$ is a natural limiting case, in which $\phi$ has zero scaling dimension and $\Delta(\tau,x)$  acquires the $\log\tau^2 m^2$ term.

\subsection{The scalar in a curved HLG background}
The $z=3$ Lifshitz scalar has a natural coupling to $z=3$ HLG; moreover, it is easy to construct a Weyl-invariant coupling.
To describe this, we will first review some basic features of $z=3$ HLG.  We follow~\cite{Horava:2009uw}. 

The HLG theory is defined on a four-dimensional space-time equipped with a co-dimension one foliation, where the latter structure encodes the privileged role of the time direction.  The degrees of freedom are familiar from the ADM decomposition in Einstein gravity:  there is a metric on spatial slices, $g_{ij}(\tau,x)$, a shift one-form $N_i (\tau,x)$, and a lapse function $\nu(\tau,x)$.  The action is given by
\begin{equation}
S_{\text{HLG}} =  \int d\tau d^3 x ~\sqrt{g} \nu \left[\frac{2}{\kappa^2}(K_{ij} K^{ij} -\lambda K^2)+\frac{\kappa^2}{2w^4} C_{ij} C^{ij} \right],
\end{equation}
where $\kappa,w$ and $\lambda$ are undetermined parameters, $C_{ij}$ is the Cotton tensor constructed from $g_{ij}$, and 
\begin{equation}
K_{ij} = \dot{g}_{ij} - \nabla_i N_j - \nabla_j N_i
\end{equation}
is the extrinsic curvature. The connection $\nabla$ is the Levi-Civita connection associated to $g$.  

$S_{\text{HLG}}$ is invariant under spatial diffeomorphisms, time reparametrizations, and for $\lambda = 1/3$ under local Weyl rescaling.  Denoting the infinitesimal parameters for these transformations by $\xi^i(\tau,x)$, $f(\tau)$, and $\omega(\tau,x)$, the action on the fields is
\begin{eqnarray}
\label{eq:algebra}
\delta g & =& \cL_{\xi} g + f\p_\tau g + 2\omega g, \nonumber\\
\delta N & = & \cL_{\xi} N +\p_\tau(f N) + \p_\tau \xi \llcorner g + 2\omega N, \nonumber\\
\delta\nu & = & \cL_{\xi} \nu + \p_\tau(f \nu) + 3 \omega \nu.
\end{eqnarray}
Here $\cL_{\xi}$ denotes the Lie derivative with respect to the vector field $\xi$, and $\llcorner$ denotes the interior product, viz.\ $(\xi \llcorner g)_i = \xi^j g_{ji}$.  It is a simple matter to show that the algebra closes.

Having described the gravitational theory we have in mind, we now turn back to the $z=3$ scalar.  An action in the curved background that reduces to the correct flat space limit and has diffeomorphism and time reparametrization invariances is given by
\begin{equation}
S = \int d\tau d^3 x~\sqrt{g}\left\{\frac{1}{2\nu} (\dot{\phi} -g^{ij} N_i \p_j \phi)^2 - \frac{\nu}{2} \phi (\nabla^2)^3 \phi \right\}.
\end{equation}
$S$ is invariant, provided $\phi$ transforms as a scalar:
\begin{equation}
\delta\phi = \xi^i \p_i \phi + f\dot{ \phi}.
\end{equation}
In fact, the kinetic term is also Weyl invariant if $\phi$ has (as expected) Weyl weight zero.  So, to construct a Weyl-invariant action we just need to make a suitable modification of the potential term.  This is easily achieved by introducing a Weyl-covariant derivative $\nablat$. 

To describe the construction, let $\nablah$ denote the Levi-Civita connection associated to the Weyl-invariant metric $\gh = \nu^{-2/3} g$.  Given a tensor $T$ with Weyl weight $q$, we define $\nablat T$ via
\begin{equation}
\nablat T = \nablah T - q A \otimes T ,
\end{equation}
where $A = \frac{1}{3} d \log \nu$.  It is easy to see that under a Weyl transformation $\delta_\omega \nablat T = q \omega \nablat T$.  
Furthermore, since $d  A = 0$, the curvature associated to $\nablat$ is just the Riemann curvature associated to the metric $\gh$.  Finally,
two useful properties of $\nablat$ are
\begin{equation}
\nablat \nu = 0,~~~ \nablat g = 0.
\end{equation}

Using the connection $\nablat$ we can construct a Weyl-invariant potential term.  In fact, many choices are possible.  In what follows, we will restrict our considerations to a two-parameter family of terms, replacing $\nu \phi (\nabla^2)^3 \phi$ with
\begin{equation}
S_V  =  \frac{1}{2} \int d\tau d^3x ~\sqrt{g} \nu \left[ \alpha \nablat_i \nablat_j \nablat_k \phi \nablat^i \nablat^j \nablat^k \phi
+\beta \nablat^2 \nablat_k \phi \nablat^2 \nablat^k \phi + \gamma \nablat_k \nablat^2 \phi \nablat^k \nablat^2 \phi \right],
\end{equation}
with non-negative parameters $\alpha,\beta,\gamma$ constrained by $\alpha+\beta+\gamma =1$.  Note that we have chosen these terms to keep the improved action positive-definite.  Such a requirement, while not very sensible for theories coupled to Einstein gravity, does seem to make sense in the context of HLG, since the latter minimal action is itself positive-definite.

\subsection{Conservation laws in flat space}
Having derived the curved-space action, we can follow the usual logic to find the conservation laws in the flat background (i.e. $N=0,\nu=1$, $g_{ij} = \delta_{ij}$).  To this end, we compute
\begin{equation}
\delta S |_{\text{flat}} = \int d\tau d^3 x~ \left[ -\delta\nu \cE -\frac{1}{2} \delta g_{il} T^{il} - \delta N_i P^i \right],
\end{equation}
with\footnote{We will use a short-hand:  $\phi_{i_1\cdots i_n} \equiv \p_{i_1}\cdots \p_{i_n} \phi$.}
\begin{equation}
P_i = \dot{\phi} \phi_i,
\end{equation}
\begin{eqnarray}
\cE & = & \ff{1}{2} \dot{\phi}^2 + \ff{5\alpha}{6} \phi_{ijk}^2 +\ff{\beta-4\alpha-3\gamma}{6} (\p^2\phi_i)^2 +\ff{\beta-\alpha-\gamma}{3} \phi_i (\p^2)^2\phi_i \nonumber\\
~&~&~~+ \ff{\alpha+3\beta+2\gamma}{3}\p^2\phi (\p^2)^2\phi -\gamma \phi_{jk} \p^2\phi_{jk},
\end{eqnarray}
and
\begin{eqnarray}
T_{il} & = & -\delta_{il} \left[\ff{1}{2}\dot{\phi}^2 + \ff{\alpha}{2} \phi_{jkm}^2 +\ff{\beta-\gamma}{2} (\p^2\phi_k)^2
+\beta \phi_k (\p^2)^2 \phi_k +\beta \p^2\phi (\p^2)^2\phi - \gamma \phi_{jk} \p^2 \phi_{jk} \right] \nonumber\\
~&~&-\alpha\phi_{ijk} \phi_{ljk} + (\beta+\gamma) \p^2\phi_i \p^2\phi_l + (2\alpha-\gamma)\phi_{ilk}\p^2\phi_k + 2\alpha \phi_{iljk} \phi_{jk} 
         -\gamma\phi_{il} (\p^2)^2\phi \nonumber\\
~&~&-(\alpha+\gamma) \p^2\phi_{il} \p^2\phi + \phi_l (\p^2)^2\phi_i + \phi_i (\p^2)^2\phi_l -\alpha(\phi_{ik}\p^2\phi_{lk} +\phi_{lk}\p^2\phi_{ik}) \nonumber\\
~&~&-(\alpha+\gamma) \p^2\phi_{ilk}\phi_k.
\end{eqnarray}

By using the variations in (\ref{eq:algebra}), we extract the following conservation laws.  Time reparametrization invariance leads to
\begin{equation}
\dot{E} \equiv \int d^3 x \p_\tau \cE = 0;
\end{equation}
spatial diffeomorphism invariance leads to
\begin{equation}
J_m \equiv \p_\tau P_m + \p_{n} T_{nm} = 0;
\end{equation}
and, finally, Weyl invariance implies
\begin{equation}
W \equiv 3 \cE + T_{ii} = 0.
\end{equation}
A short computation shows $W=0$ identically, while $\dot{E} = 0$ and $J_m=0$ hold modulo the equation of motion. 

\subsection{Quantum corrections and the anomaly}
Having described the classical conservation laws in flat space, we have two possible routes to determining the form of the anomaly.  The first, which would produce the most satisfactory results, would be to construct the renormalized quantities $P_i$, $T_{ij}$ and $\cE$ in an arbitrary gravitational background.  This sort  of computation is already quite heroic in relativistic theories~\cite{Birrell:1982ix} and would be  challenging to implement here.  The second, which is how the relativistic conformal anomaly was discovered in the first place~\cite{Duff:1993wm}, is to study the $n$-point functions of $P_i$, $T_{ij}$ and $\cE$ in flat space and extract the presence of the anomaly from these.  It is this second method we will pursue below.

In order to preserve the conservation laws $J_m = 0$, $\dot{E}=0$, and $W=0$ we must find local counter-terms so that, for example,
\begin{equation}
\la (\dot{P}_i + \p_j T_{ji} )(x) \cE(0) \ra + \p_\tau \la P_i(x) \cE(0)\ra_{\text{ct}} + \p_j \la T_{ji}(x) \cE(0)\ra_{\text{ct}} = 0.
\end{equation}

In Lorentz-invariant theories this would be  a simple one-loop computation, easily performed in either momentum or position space.  In case of a Lifshitz theory, the problem is more involved.  Some methodology is available for handling loop integrals in Lifshitz theories~\cite{Carvalho:2009nv}, but the expressions for even simple diagrams with non-zero external momenta are quite forbidding.  The expressions simplify for one-loop graphs with zero external momenta, leading to, for instance, tractable computations of beta functions and gap equations~\cite{Dhar:2009dx,Das:2009ba,Iengo:2009ix}.  To compute correlators at non-zero external momenta, we found it easier to consider the computation in position space.  Setting up this computation in a convenient regularization scheme will occupy us in what follows.


\section{A regularization and position space
  computations} \label{s:regularization}

In this section we present a regularization scheme and apply  it to anomaly computations for relativistic scalars, where
the results are well-known.

\subsection{The regularization scheme} \label{sec:regularization-scheme}
In the course of position-space computations in a relativistic, $d$-dimensional, scale-invariant theory, one typically 
encounters singular distributions such as $|x|^{-2d}$. A nice way to deal with these distributions is via the method of differential 
regularization~\cite{Freedman:1991tk,Osborn:1993cr}.  However, since our final goal is to study the anisotropic theory, we will instead
introduce a smeared propagator as was considered in~\cite{Freedman:1992gr}.  That is, we introduce a scale $\ep$ and replace the 
standard propagator $\Delta(x^2)$ with
\begin{equation}
\Delta_\ep (x^2) = \Delta(x^2 + \ep^2).
\end{equation}
This leads to a representation of the Dirac $\delta$-function via Green's equation,
\begin{equation}
\delta_\ep(x) \equiv -\p^2 \Delta_\ep(x^2) \ ,
\end{equation}
since
\begin{equation}
\lim_{\ep \to 0} \int d^d x ~\delta_\ep(x) f(x) = f(0)
\end{equation}
for any bounded function $f(x)$.

In our computations we will encounter distributions $D_\ep (x)$ that satisfy two basic properties: $D_\ep(x)$ is a smooth bounded function with $\lim_{\ep\to 0} D_\ep(x) = 0$ for any $x\neq 0$; and $D_\ep(x) \sim \ep^k |x|^{-n}$ for large $|x|$, with $k>0$ and $n>d$.  We can find a
convenient representation for such distributions by integrating them against a smooth, bounded test function $f(x)$.  Specifically, we have
\begin{eqnarray}
\int d^d x~ D_\ep(x) f(x) &=& \int d^d y~ D_{1}(y) \ep^{k-n+d} f(\ep y) \nonumber\\
~&=& \int d^d y D_{1} (y) \ep^{k-n+d} 
\left[\sum_{m=0}^{n-d-k} \frac{\ep^m}{m!} y^{i_1}\cdots y^{i_m} \p_{i_1\cdots i_m} f(0) \right]\nonumber\\
~&~&+\int d^d y D_{1} (y) \ep^{k-n+d} \left[f(\ep y) -
  \sum_{m=0}^{n-d-k} \frac{\ep^m}{m!} y^{i_1}\cdots y^{i_m}
  \p_{i_1\cdots i_m} f(0) \right] \ .\nonumber\\
\end{eqnarray}
Since $f(x)$ is smooth and bounded and $\int d^d y D_1(y) |y|^m < \infty$ for $m< n-d$, it follows that the second line is a convergent integral for $\ep\neq 0$.  Furthermore, by Taylor's theorem the integrand in the second line scales as $\ep$ for small $\ep$, so that this remainder term vanishes in the $\ep \to 0$ limit.  Keeping the terms that do not vanish as $\ep \to 0$, we obtain
a representation for the distribution:
\begin{equation}
D_\ep(x) \to \sum_{m=0}^{n-d-k} \frac{\ep^{m+k+d-n}(-)^m}{m!} S^{i_1\cdots i_m} \p_{i_1\cdots i_m} \delta^{(d)}(x),
\end{equation}
where the coefficients $S^{i_1\cdots i_m}$ are obtained by computing the convergent integrals
\begin{equation}
S^{i_1\cdots i_m} = \int d^d y~ D_1(y) y^{i_1}\cdots y^{i_m}.
\end{equation}

This regularization may be adapted to the Lifshitz scalar with two noteworthy modifications.  First, a look at the $z=3$ propagator given in (\ref{eq:prop}) shows that it is sufficient to do the smearing in the time direction.  That is, we will replace
\begin{equation} \label{eq:regularized-HL-propagaor}
\Delta(\tau^2, x^2) \to \Delta_\ep(\tau^2, x^2) = \Delta(\tau^2+\ep^6, x^2).
\end{equation}
Second, while integrating the distributions we will encounter against a smooth bounded test function still leads to sensible representations, the integrals that must be evaluated are quite a bit more difficult, involving a large number of terms of products of hypergeometric functions.  We have opted to handle these numerically.

\subsection{The relativistic scalar in two
  dimensions} \label{sec:relat-scal-two-dim}

As a warm up exercise we treat the well-known case of the relativistic
scalar in two dimensions using the prescription of $\epsilon$
regularization introduced above.

The action of a free scalar in two dimensions is
\begin{equation}
  S = \int d^2 x \, \partial_\mu \phi \partial^\mu \phi \ ,
\end{equation}
where $\mu, \nu, \dots = 1, 2$ and we work in a Euclidean signature so
indices are raised and lowered using the metric $\delta_{\mu
  \nu}$. The regularized scalar propagator is
\begin{equation}
  \Delta_\epsilon (x) = -\frac{1}{4 \pi} \log (x^2 + \epsilon^2) \ ,
\end{equation}
leading to a representation for the Dirac $\delta$-function
\begin{equation}
  \delta_\epsilon = \frac{\epsilon^2}{\pi (x^2 + \epsilon^2) ^2} \ .
\end{equation}
The energy-momentum tensor is
\begin{equation}
  T_{\mu \nu} = \partial_\mu \phi \partial_\nu \phi - \frac{1}{2}
  \delta_{\mu \nu} \partial_\lambda \phi \partial^\lambda \phi \ .
\end{equation}
This tensor is conserved up to the equations of motion and is
identically traceless.

The first step in the analysis of the anomaly is to check whether the
symmetry under diffeomorphisms is violated quantum mechanically in the
presence of a gravitational background. This is captured to first
order in the metric perturbation by the two-point function
\begin{equation}
  C_{\nu \alpha \beta} = \langle \partial^\mu T_{\mu \nu} (x) T_{\alpha
    \beta} (0) \rangle \ .
\end{equation}
Performing the Wick contractions and using the regularized propagator
we have
\begin{eqnarray}
  C_{\nu \alpha \beta} & = & \partial_\alpha \partial^2 \Delta_\epsilon (x)
  \partial_{\beta \nu} \Delta_\epsilon (x) + \partial_\beta \partial^2
  \Delta_\epsilon (x) \partial_{\alpha \nu} \Delta_\epsilon (x) -
  \delta_{\alpha \beta} \partial^\lambda \partial^2 \Delta_\epsilon (x)
  \partial_{\nu \lambda} \Delta_\epsilon (x) = \nonumber \\
  & = & -\frac{2 \epsilon^2}{\pi^2 (x^2 +
    \epsilon^2)^4} \left[ x_\alpha \delta_{\beta \nu} + x_\beta
    \delta_{\alpha \nu} + \frac{x^2 - \epsilon^2}{x^2 + \epsilon^2}
    \delta_{\alpha \beta} x_\nu - \frac{4}{x^2 + \epsilon^2} x_\alpha
    x_\beta x_\nu \right] \ .
\end{eqnarray}
Note that $\lim_{\epsilon \to 0} C_{\nu \alpha \beta}|_{x \neq 0} =
0$, so as expected, the violation of the energy-momentum conservation is
local. In order to extract the local contact-terms the procedure
outlined in subsection \ref{sec:regularization-scheme} is used. One
then has
\begin{eqnarray}
  \int d^2 x \, C_{\nu \alpha \beta} (x) f(x) & = & -\frac{2}{\pi^2}
  \int d^2 y \frac{1}{\epsilon^3 (y^2 + 1)^4} \bigg[ y_\alpha \delta_{\beta
      \nu} + y_\beta \delta_{\alpha \nu} + \frac{y^2 - 1}{y^2
      + 1} \delta_{\alpha \beta} y_\nu - \nonumber \\
    && {} - \frac{4}{y^2 + 1} y_\alpha y_\beta y_\nu \bigg]
  f(\epsilon y) \ ,
\end{eqnarray}
where we have changed the integration variable to $x = \epsilon
y$. Expanding $f(\epsilon y)$ in powers of $\epsilon$ and doing the
angular integration (keeping in mind that terms with odd powers of $y$
vanish in the integration due to the spherical symmetry) we finally
get
\begin{eqnarray}
  C_{\nu \alpha \beta} & = & \left[ \frac{1}{6 \pi \epsilon^2} P_{\mu \nu \alpha
    \beta} \partial^\mu + \frac{1}{48 \pi} \left( \delta_{\mu \alpha}
    \delta_{\nu \beta} + \delta_{\mu \beta} \delta_{\nu \alpha} \right)
    \partial^\mu \partial^2 - \frac{1}{24 \pi} \partial_{\nu \alpha
      \beta} \right] \delta^{(2)} (x) \ , \nonumber \\
  P_{\mu \nu \alpha \beta} & = & \frac{1}{2} ( \delta_{\mu \alpha}
  \delta_{\nu \beta} + \delta_{\mu \beta} \delta_{\nu \alpha}
  -\delta_{\mu \nu} \delta_{\alpha \beta}) \ ,
\end{eqnarray}
where only the terms which do not vanish in the limit $\epsilon \to 0$
have been kept. Hence, we see that energy-momentum conservation is
violated quantum mechanically.

In order to preserve energy-momentum conservation (and the related
symmetry under diffeomorphism) we introduce the following
local counter-term
\begin{equation}
  \langle T_{\mu \nu} (x) T_{\alpha \beta} (0) \rangle_\mathrm{ct} =
  \left[ b P_{\mu \nu \alpha \beta} + a (\delta_{\beta \nu} \tilde
    \partial_\mu \tilde \partial_\alpha + \delta_{\beta \mu} \tilde
    \partial_\nu \tilde \partial_\alpha + \delta_{\alpha \nu} \tilde
    \partial_\mu \tilde \partial_\beta + \delta_{\alpha \mu} \tilde
    \partial_\nu \tilde \partial_\beta) \right] \delta^{(2)}(x) \ ,
\end{equation}
where $\tilde \partial_\mu = \epsilon_{\mu \nu} \partial^\nu$.
Restoration of energy-momentum conservation requires that
\begin{equation}
  a = -\frac{1}{48 \pi} \ , \quad b = -\frac{1}{6 \pi \epsilon^2} \ .
\end{equation}
However, this clashes with the Weyl symmetry since
\begin{equation}
  \langle T_\mu^\mu (x) T_{\alpha \beta} (0) \rangle_\mathrm{ct} =
  -\frac{1}{12 \pi}
  \tilde \partial_\alpha \tilde \partial_\beta \delta^{(2)}(x) \ .
\end{equation}
Hence, the regularization scheme used here reproduces the familiar Weyl
anomaly (e.g.\ in \cite{Green:1987sp}).

Finally, in order to determine the type of the anomaly we follow
\cite{Osborn:1993cr} and compute the dependence of the two-point
function on the regularization scale $\epsilon$
\begin{equation} \label{eq:two-point-log-derivative}
  \epsilon \frac{\partial}{\partial \epsilon} \langle T_{\mu \nu}(x)
  T_{\rho \sigma}(0) \rangle \ .
\end{equation}
Treating the resulting distribution using the regularization scheme, we
find
\begin{eqnarray}
  \epsilon \frac{\partial}{\partial \epsilon} \langle T_{\mu \nu}(x)
  T_{\rho \sigma}(0) \rangle & = & \frac{1}{6 \pi} \bigg[
    -\frac{2}{\epsilon^2} P_{\mu \nu \rho \sigma} - \nonumber \\
    && {} - \delta_{\mu \nu} \delta_{\rho \sigma} \partial^2 +
    \frac{1}{2} (\delta_{\mu \rho}  \delta_{\nu \sigma} + \delta_{\mu
      \sigma} \delta_{\nu \rho}) \partial^2 + \nonumber \\
    && {} + I_{\mu \nu \rho \sigma} - \frac{1}{2}
    (I_{\mu \rho \nu \sigma} + I_{\mu \sigma \nu \rho}) \bigg]
  \delta^{(2)}(x) + O(\epsilon) \ , \nonumber \\
  I_{\mu \nu \rho \sigma} & \equiv & \delta_{\mu \nu} \partial_{\rho
    \sigma} + \delta_{\rho \sigma} \partial_{\mu \nu} \ .
\end{eqnarray}
At first sight, it seems as if the logarithmic derivative has a finite
term and a scale $\mu$ needs to be introduced to properly define
$\langle T_{\mu \nu} T_{\rho \sigma} \rangle$. However, the indices in
two dimensions can take only two values, so the finite term actually
vanishes and
\begin{equation}
  \epsilon \frac{\partial}{\partial \epsilon} (\langle T_{\mu \nu}
  (x) T_{\rho \sigma} (0) \rangle + \langle T_{\mu \nu} (x) T_{\rho
    \sigma} (0) \rangle_\mathrm{ct}) = 0\ .
\end{equation}
As expected, the anomaly is type A in two dimensions.

\subsection{The relativistic scalar in four dimensions}

In order to demonstrate the regularization method applied to a known
case when there is a type B Weyl anomaly, we consider a scalar field
coupled conformally to gravity. The regulated propagator in this case
is
\begin{equation}
  \Delta_\epsilon (x) = \frac{1}{4 \pi^2} \frac{1}{x^2 + \epsilon^2}
  \ ,
\end{equation}
and the regularized version of the $\delta$-function satisfying
$\partial^2 \Delta_\epsilon (x) = -\delta_\epsilon(x)$ is given by
\begin{equation}
  \delta_\epsilon (x) = \frac{2 \epsilon^2}{\pi^2 (x^2 + \epsilon^2)^3} \ .
\end{equation}
The improved energy-momentum tensor is \cite{Osborn:1993cr}
\begin{equation}
  T_{\mu \nu} = \partial_\mu \phi \partial_\nu \phi - \frac{1}{12}
  (\delta_{\mu \nu} \partial^2 + 2 \partial_\mu \partial_\nu) \phi^2
  \ .
\end{equation}

The first thing to be done is to compute the violation of the
energy-momentum conservation in the two-point function. Using the same
techniques as in the $d=2$ example, we find
\begin{eqnarray}
  \lefteqn{\langle \partial^\mu T_{\mu \nu}(x) T_{\rho \sigma}(0) \rangle =
  \frac{1}{60 \pi^2}\bigg\{ \frac{1}{\epsilon^4} \Big[ A_1
    \delta_{\rho \sigma} \partial_\nu + 
    A_2 (\delta_{\nu \rho} \partial_\sigma + \delta_{\nu \sigma}
    \partial_\rho) \Big] +} \nonumber \\
  && {} + \frac{1}{\epsilon^2} \Big[ A_3 \partial_{\nu \rho \sigma} +
    A_4 \delta_{\rho \sigma} \partial_\nu \partial^2 
    + A_5 (\delta_{\nu \rho} \partial_\sigma + \delta_{\nu \sigma}
    \partial_\rho) \partial^2 \Big] + \nonumber \\
  && {} + \left[ A_6 \partial_{\rho \sigma \nu} + A_7
  \delta_{\rho \sigma} \partial_\nu \partial^2 + A_8 (\delta_{\nu
    \rho} \partial_\sigma + \delta_{\nu \sigma} \partial_\rho)
  \partial^2 \right] \partial^2 \bigg\} \delta^{(4)} (x)
  \ , \label{eq:4d-relativistic-conservation-two-pt}
\end{eqnarray}
where the coefficients are
\begin{eqnarray}
  A_1 & = & A_2 = 1 \ , \quad
  A_3 = -\frac{1}{6} \ , \quad
  A_4 = -\frac{1}{12} \ , \nonumber \\
  A_5 & = & \frac{1}{8} \ , \quad
  A_6 = -\frac{1}{24} \ , \quad
  A_7 = -\frac{7}{192} \ , \quad
  A_8 = \frac{1}{64} \ .
\end{eqnarray}

Next, we turn to the violation of the Weyl symmetry in the two-point
function. Here the four-dimensional theory differs from the
two-dimensional one since the trace only vanishes up to the equation
of motion. The most general possible contact term allowed by Lorentz
and Bose symmetries and dimensional arguments is
\begin{eqnarray}
  \langle T_\mu^\mu(x) T_{\rho \sigma}(0) \rangle & = & \frac{1}{60 \pi^2}
  \bigg[ \frac{1}{\epsilon^4} B_1 \delta_{\rho \sigma} +
    \frac{1}{\epsilon^2} (B_2 \delta_{\rho \sigma} \partial^2 +B_3
    \partial_\rho \partial_\sigma) + (B_4 \delta_{\rho \sigma}
    \partial^2 + \nonumber \\
    && {} + B_5 \partial_\rho \partial_\sigma) \partial^2 \bigg]
  \delta^{(4)}(x) \ . \label{eq:4d-relativitsic-trace-two-pt}
\end{eqnarray}
Direct computation leads to the coefficients
\begin{equation}
  B_1 = 6 \ , \quad
  B_2 = -\frac{1}{12} \ , \quad
  B_3 = -\frac{1}{6} \ , \quad
  B_4 = -\frac{11}{96} \ , \quad
  B_5 = -\frac{1}{24} \ .
\end{equation}

On the other hand, the most general local counter-term consistent with
Lorentz and Bose symmetries and dimensional analysis is
\begin{eqnarray}
  \langle T_{\mu \nu} (x) T_{\rho \sigma} (0) \rangle_\mathrm{ct} & =
  & \frac{1}{60 \pi^2} \bigg\{ \frac{1}{\epsilon^4} \left[ C_1 \delta_{\mu \nu}
    \delta_{\rho \sigma} + C_2 (\delta_{\mu \rho} \delta_{\nu
      \sigma} + \delta_{\mu \sigma} \delta_{\nu \rho}) \right] +
  \nonumber \\
  && {} + \frac{1}{\epsilon^2} \big[ C_3 I_{\mu \nu \rho \sigma} 
    + C_4 (I_{\mu \rho \nu \sigma} + I_{\nu \rho \mu \sigma}) +C_5
    \delta_{\mu \nu} \delta_{\rho \sigma} \partial^2 + \nonumber \\
    && {} + C_6 (\delta_{\mu \rho} \delta_{\nu \sigma} +\delta_{\mu
      \sigma} \delta_{\nu \rho}) \partial^2 \big] + C_7 \delta_{\mu
    \nu} \delta_{\rho \sigma} (\partial^2)^2 + \nonumber \\ 
  && {} + C_8 (\delta_{\mu \rho} \delta_{\nu \sigma} + \delta_{\mu
    \sigma} \delta_{\nu \rho}) (\partial^2)^2 + C_9 I_{\mu \nu \rho
    \sigma} \partial^2 + C_{10} (I_{\mu \rho \nu \sigma} + I_{\mu
    \sigma \nu \rho}) \partial^2 + \nonumber \\ 
  && {} + C_{11} \partial_{\mu \nu \rho \sigma} \bigg \} \delta^{(4)} (x) \ . 
\end{eqnarray}
By requiring these counter-terms to cancel the contact terms in
(\ref{eq:4d-relativistic-conservation-two-pt}) and
(\ref{eq:4d-relativitsic-trace-two-pt}), we get a one-parameter family
of solutions:
\begin{eqnarray}
  C_1 & = & C_2 = -1 \ , \quad
  C_3 = -\frac{1}{12} \ , \quad
  C_4 = \frac{1}{8} \ , \quad
  C_5 = \frac{1}{6} \ , \nonumber \\
  C_6 & = & -\frac{1}{4} \ , \quad
  C_7 = \frac{11}{192} - \frac{C_{11}}{2} \ , \quad
  C_8 = -\frac{3}{64} + \frac{3 C_{11}}{4} \ , \nonumber \\
  C_9 & = & -\frac{1}{48} + \frac{C_{11}}{2} \ , \quad
  C_{10} = \frac{1}{32} - \frac{3 C_{11}}{4} \ .
\end{eqnarray}
So after adding the counter-terms the energy-momentum tensor is
conserved and traceless in the two-point function. This matches the
well-known result that the trace is non-vanishing only in the
three-point function.

However, by considering the logarithmic derivative
\begin{eqnarray}
  \epsilon \frac{\partial}{\partial \epsilon} \langle T_{\mu \nu} (x)
  T_{\rho \sigma} (0) \rangle & = & \frac{1}{60 \pi^2} \bigg\{
  \frac{1}{\epsilon^4} \big[ D_1 \delta_{\mu \nu} \delta_{\rho \sigma}
    + D_2 (\delta_{\mu \rho} \delta_{\nu \sigma} + \delta_{\mu \sigma}
    \delta_{\nu \rho}) \big] + \frac{1}{\epsilon^2} \big[ D_3
    \delta_{\mu \nu} \delta_{\rho \sigma} \partial^2 + \nonumber \\
    && {} + D_4 (\delta_{\mu \rho} \delta_{\nu \sigma} + \delta_{\mu
      \sigma} \delta_{\nu \rho}) \partial^2 + D_5 I_{\mu \nu \rho
      \sigma} + D_6 (I_{\mu \rho \nu \sigma} + I_{\nu \rho \mu
      \sigma}) \big] + \nonumber \\
    && {} + D_7 \delta_{\mu \nu} \delta_{\rho
      \sigma} (\partial^2)^2 + D_8 (\delta_{\mu \rho} \delta_{\nu
      \sigma} + \delta_{\mu \sigma} \delta_{\nu \rho}) (\partial^2)^2
    + D_9 I_{\mu \nu \rho \sigma} \partial^2 + \nonumber \\
    && {} + D_{10} (I_{\mu \rho \nu \sigma} + I_{\nu \rho \mu \sigma}) \partial^2
    + D_{11} \partial_{\mu \nu \rho \sigma} \bigg\} \delta^{(4)}(x)
\end{eqnarray}
it is possible to determine whether a scale signaling the violation of
the Weyl symmetry is introduced. Direct computation of the
coefficients using the regularization and the techniques presented
above yields
\begin{eqnarray}
  D_1 & = & D_2 = -4 \ , \quad
  D_3 = \frac{1}{3} \ , \quad
  D_4 = -\frac{1}{2} \ , \quad
  D_5 = -\frac{1}{6} \ , \quad
  D_6 = \frac{1}{4} \ , \nonumber \\
  D_7 & = & \frac{1}{24} \ , \quad
  D_8 = -\frac{1}{16} \ , \quad
  D_9 = -\frac{1}{24} \ , \;\;
  D_{10} = \frac{1}{16} \ , \quad
  D_{11} = -\frac{1}{12} \ .
\end{eqnarray}
The logarithmic derivative of the renormalized energy-momentum
two-point function (which includes the contribution for the
counter-terms) is
\begin{eqnarray}
  \lefteqn{\epsilon \frac{\partial}{\partial \epsilon} (\langle T_{\mu \nu} (x)
  T_{\rho \sigma} (0) \rangle + \langle T_{\mu \nu} (x) T_{\rho
    \sigma} (0) \rangle_\mathrm{ct}) =   \frac{1}{2880 \pi^2} \big[
    2 \delta_{\mu \nu} \delta_{\rho \sigma} (\partial^2)^2 -} \nonumber \\
    && {} - 3(\delta_{\mu \rho} \delta_{\nu \sigma} + \delta_{\mu \sigma}
    \delta_{\nu \rho}) (\partial^2)^2 - 2 I_{\mu \nu \rho \sigma}
    \partial^2 + 3 (I_{\mu \rho \nu \sigma} + I_{\nu \rho \mu
      \sigma}) \partial^2 - \nonumber \\
    && {} - 4 \partial_{\mu \nu \rho \sigma} \big] \delta^{(4)} (x) \ .
\end{eqnarray}
We see that the terms which diverge as $\epsilon \to 0$ are canceled
by the counter-terms and only the finite piece remains. This result
matches the one in \cite{Osborn:1993cr} up to a minus sign coming from
the derivative being with respect to a length scale and not a mass
scale --- lending credence to our regularization method. Integrating
the above equation leads to a logarithmic term $\log(\mu \epsilon)$
with $\mu$ being a scale introduced in order to make the argument of
the logarithm dimensionless. Thus the renormalized two-point function
and hence the effective action contain a scale $\mu$. This signals
that the Weyl symmetry is broken by a type B anomaly.

\section{The Weyl anomaly}\label{s:anomaly}

In this section we show that the conformally-coupled scalar in a Ho\v
rava-Lifshitz background induces a Weyl anomaly.

The propagator of a scalar coupled conformally to a flat Ho\v
rava-Lifshitz background (\ref{eq:regularized-HL-propagaor}), together
with the Green's equation $\Box \Delta_\epsilon (\tau, x) = -
\delta_\epsilon (\tau, x)$, where $\Box \equiv \partial_\tau^2 +
(\partial^2)^3$, leads to the regularized $\delta$-function
\begin{eqnarray}
  \delta_\epsilon (\tau, x) & = & \frac{\epsilon^6}{18 \pi^2 (\tau^2 +
    \epsilon^6)^2} \left( u_\epsilon \frac{\partial}{\partial
    u_\epsilon} + 3 \right) F (u_\epsilon) \ ,\nonumber \\
  F (u) & \equiv & \sum_{n = 0}^\infty \frac{\Gamma \left(
    \frac{2 n}{3} + 1 \right)}{(2n + 1)!} (-u)^n
  \ ,  \label{eq:HL-Dirac-delta} \,
\end{eqnarray}
where $u_\epsilon = \frac{x^2}{(\tau^2 + \epsilon^6)^{1/3}}$. We
demonstrate in Appendix \ref{sec:useful-formulas} that 
\begin{equation}
  \int d\tau d^3 x \, \delta_\epsilon (\tau, x) = 1 \ .
\end{equation}

\subsection{Counter-term analysis}

As in the relativistic examples, we begin with the quantum violation of
energy and momentum conservation Ward identities. As expected from
dimensional analysis, rotational invariance, parity and time reversal
symmetry, they are of the form
\begin{equation}
  \langle \dot E(\tau) \cE(0,0) \rangle = \left(
  \frac{1}{\epsilon^6} A_{\cE \cE 1} + A_{\cE \cE 2} \partial_\tau^2
  \right) \partial_\tau \delta (\tau) \ ,
\end{equation}
\begin{equation}
  \langle \dot E(\tau) T^{i j} (0, 0) \rangle =
  \delta^{ij} \left( \frac{1}{\epsilon^6} A_{\cE T 1} + A_{\cE T 2}
  \partial_\tau^2 \right) \partial_\tau \delta (\tau) \ ,
\end{equation}
\begin{equation}
  \langle \dot E(\tau) P^i (0, 0) \rangle = 0 \ ,
\end{equation}
\begin{equation}
  \langle J^i (\tau, x) P^j (0,0) \rangle = \left[
    \frac{1}{\epsilon^2} A_{JP1} \delta^{ij} \partial_\tau + (A_{JP2}
    \partial^{ij} + A_{JP3} \delta^{ij} \partial^2) \partial_\tau
    \right] \delta^{(4)}(\tau, x) \ , \label{eq:JP-two-point-function}
\end{equation}
\begin{eqnarray}
  \langle J^i (\tau, x) T^{lm} (0, 0) \rangle & = & \bigg\{
  (\delta^{mi} \delta^{kl} + 
  \delta^{il} \delta^{mk}) \partial_k \bigg[ \frac{1}{\epsilon^6}
    A_{JT1} + \frac{1}{\epsilon^4} A_{JT2} \partial^2 +
    \frac{1}{\epsilon^2} A_{JT3} (\partial^2)^2 + \nonumber \\
    && {} + A_{JT4} (\partial^2)^3 +
    A_{JT5} \partial_\tau^2 \bigg] + \delta^{lm} \partial^i \bigg[
    \frac{1}{\epsilon^6} A_{JT6} + \nonumber \\
    && {} + \frac{1}{\epsilon^4} A_{JT7} \partial^2 + \frac{1}{\epsilon^2}
    A_{JT8} (\partial^2)^2 + A_{JT9} (\partial^2)^3 + \nonumber \\
    && {} + A_{JT10} \partial_\tau^2 \bigg] 
  + \partial^{ilm}
  \bigg[ \frac{1}{\epsilon^4} A_{JT11} + \frac{1}{\epsilon^2} A_{JT12}
    \partial^2 + \nonumber \\
    && {} + A_{JT13} (\partial^2)^2 \bigg] \bigg\} \delta^{(4)} (\tau,
  x) \ ,
\end{eqnarray}
\begin{eqnarray}
  \langle J^m (\tau, x) \cE (0, 0) \rangle & = & \partial^m \bigg[
    \frac{1}{\epsilon^6} A_{J \cE 1} + \frac{1}{\epsilon^4} A_{J \cE
      2} \partial^2 + \frac{1}{\epsilon^2} A_{J \cE 3} (\partial^2)^2
    + A_{J \cE 4} (\partial^2)^3 + \nonumber \\
    && {} + A_{J \cE 5} \partial_\tau^2 \bigg] \delta^{(4)} (\tau, x)
  \ .
\end{eqnarray}
The Weyl Ward identity is not corrected before the introduction of
counter-terms since it vanishes identically.

We now illustrate the procedure for computing the coefficients in the
contact terms by evaluating $A_{JP1}$, $A_{JP2}$ and $A_{JP3}$ in
(\ref{eq:JP-two-point-function}). The Wick contractions yield
\begin{equation}
  C^{ij} = \langle J^i(\tau, x) P^j(0, 0) \rangle = -\partial^i \partial_\tau
  \Delta_\epsilon (\tau, x) \partial^j \delta_\epsilon (\tau, x) -
  \partial^{ij} \Delta_\epsilon (\tau, x) \partial_\tau
  \delta_\epsilon (\tau, x) \ .
\end{equation}
Substituting the expressions for $\Delta_\epsilon$ and
$\delta_\epsilon$, it can be verified that this distribution meets the
requirements listed in subsection \ref{sec:regularization-scheme} and
the use of the test function approach is possible.

To extract the expression for the distribution we consider
\begin{equation}
  I = \int d\tau d^3 x C^{ri}(\tau, x) f(\tau, x)
\end{equation}
for a smooth and bounded test function $f(\tau, x)$. We first change
variables to $x^i = \epsilon u^i$, $\tau = \epsilon^3 t$, and expand
the test function $f(\epsilon u, \epsilon^3 t)$ in powers of
$\epsilon$. Performing the angular integration and changing variables
to $v = u^2 / (t^2 + 1)^{1/3}$ in order to disentangle the $u$ and
$t$ integrations, we obtain $I = I_1 + I_2$, where
\begin{eqnarray}
  I_1 & = & \frac{1}{27 \pi^3 \epsilon^2} \partial_\tau f
  \delta^{ri} \int_{-\infty}^\infty \frac{dt \, t^2}{(t^2 + 1)^{17/6}}
  \int_0^\infty du \, u^{1/2} \left( \frac{2}{3} u H_1(u) + H_2(u) \right)
  \ , \nonumber \\
  I_2 & = & \frac{1}{54 \pi^3} \partial_\tau \partial_{kl} f
  \int_{-\infty}^\infty \frac{dt \, t^2}{(t^2 + 1)^{5/2}} \left[
    \frac{2}{15} (\delta^{ri} \delta^{kl} + 2 \delta^{rk} \delta^{il})
    J_1 + \frac{1}{3} \delta^{ri} \delta^{kl} J_2 \right] \ ,
\end{eqnarray}
\begin{eqnarray}
  F_1 (u) & = & \frac{1}{3} \left( u \frac{\partial}{\partial u} + 3
  \right) F(u) \ , \nonumber \\
  H_1 (u) & = & (G' + 2 u G'') F_1' + 6 G'' F_1 \ ,
  \nonumber \\
  H_2 (u) & = & u G' F_1' + 6 G' F_1 \ ,
\end{eqnarray}
and
\begin{eqnarray}
  J_1 & = & \int_0^\infty du \, u^{5/2} H_1(u) \ , \nonumber \\
  J_2 & = & \int_0^\infty du \, u^{3/2} H_2(u) \ .
\end{eqnarray}
$J_1$ and $J_2$ have been evaluated numerically and are convincingly
given by
\begin{equation}
  J_1 = -\frac{1}{4} J_2 \ , \quad J_2 = -\frac{27}{16} \pi \ .
\end{equation}
Hence,
\begin{equation}
A_{JP1} = 2.02\times 10^{-3},\qquad A_{JP2} = -\frac{1}{720\pi^2},\qquad A_{JP3} = \frac{1}{160\pi^2}.
\end{equation}
The rest of the coefficients are determined by the same procedure. The
numeric results are given in Appendix \ref{sec:count-term-coeff}.

The most general local counter-terms with the appropriate dimension
consistent with three-dimensional rotational symmetries, parity, time
reversal and Bose symmetry are
\begin{eqnarray}
  \langle \cE \cE \rangle_\mathrm{ct} & = & \bigg[ \frac{1}{\epsilon^6} C_{\cE \cE
    1} + \frac{1}{\epsilon^4} C_{\cE \cE 2} \partial^2 +
    \frac{1}{\epsilon^2} C_{\cE \cE 3}(\partial^2)^2 + C_{\cE \cE 4} (\partial^2)^3
    + \nonumber \\
    && {} + C_{\cE \cE 5} \partial_\tau^2 \bigg] \delta^{(4)}(\tau, x) \ ,
\end{eqnarray}
\begin{eqnarray}
  \langle \cE T^{ij} \rangle_\mathrm{ct} & = & \bigg\{ \delta^{ij} \left[
    \frac{1}{\epsilon^6} C_{\cE T 1} + \frac{1}{\epsilon^4} C_{\cE T
      2} \partial^2 + \frac{1}{\epsilon^2} C_{\cE T 3} (\partial^2)^2
    + C_{\cE T 4} (\partial^2)^3 + C_{\cE T 5} \partial_\tau^2 \right]
  + \nonumber \\
  && {} + \partial^{ij} \left[ \frac{1}{\epsilon^4} C_{\cE T6}
    + \frac{1}{\epsilon^2} C_{\cE T 7} \partial^2 + C_{\cE T8}
    (\partial^2)^2 \right] \bigg\} \delta^{(4)}(\tau, x) \ ,
\end{eqnarray}
\begin{equation}
  \langle \cE P^i \rangle_\mathrm{ct} = C_{\cE P1} \partial^i
  \partial_\tau \delta^{(4)}(\tau, x)  \ ,
\end{equation}
\begin{equation}
  \langle P^i P^j \rangle_\mathrm{ct} = \left[ \delta^{ij} \left(
  \frac{1}{\epsilon^2} C_{PP1} + C_{PP2} \partial^2 \right)
  + C_{PP3} \partial^{ij} \right] \delta^{(4)}(\tau, x) \ ,
\end{equation}
\begin{equation}
  \langle P^i T^{jk} \rangle_\mathrm{ct} = \left[ \delta^{jk} C_{PT1}
  \partial^i\partial_\tau + (\delta^{ij} \partial^k + \delta^{ik}
  \partial^j) C_{PT2} \partial_\tau \right] \delta^{(4)}(\tau, x) \ ,
\end{equation}
\begin{eqnarray}
  \langle T^{ij} T^{kl} \rangle_\mathrm{ct} & = & \bigg\{ \delta^{ij}
  \delta^{kl} \bigg[ \frac{1}{\epsilon^6} C_{TT1} +
    \frac{1}{\epsilon^4} C_{TT2} \partial^2 + \frac{1}{\epsilon^2}
    C_{TT3} (\partial^2)^2 + C_{TT4} (\partial^2)^3 
    + C_{TT5} \partial_\tau^2 \bigg] + \nonumber \\
  && {} + (\delta^{ik} \delta^{jl} + \delta^{il} \delta^{jk}) \bigg[
    \frac{1}{\epsilon^6} C_{TT6} + \frac{1}{\epsilon^4} C_{TT7}
    \partial^2 + \frac{1}{\epsilon^2} C_{TT8} (\partial^2)^2 +
    \nonumber \\
    && {} + C_{TT9} (\partial^2)^3 + C_{TT10} \partial_\tau^2 \bigg] +
  I^{ijkl} \left[ \frac{1}{\epsilon^4} C_{TT11} +
    \frac{1}{\epsilon^2} C_{TT12} \partial^2 + C_{TT13} (\partial^2)^2
    \right] + \nonumber \\
  && {} + \partial^{ijkl} \left( \frac{1}{\epsilon^2} C_{TT14} +
  C_{TT15} \partial^2 \right)  + (I^{ikjl} + I^{iljk}) \bigg[
    \frac{1}{\epsilon^4} C_{TT16} + \nonumber \\
    && {} + \frac{1}{\epsilon^2} C_{TT17} \partial^2 + 
    C_{TT18} (\partial^2)^2 \bigg] \bigg\} \delta^{(4)}(\tau, x) \ .
\end{eqnarray}
The counter-terms can restore both diffeomorphism and Weyl symmetries
if the coefficients of the quantum corrections satisfy
\begin{eqnarray}
  3 A_{J \cE 5} + 3 A_{JT10} + 2 A_{JT5} & = & 0 \ , \nonumber \\
  3 A_{J \cE 4} + A_{JT13} + 2 A_{JT4} + 3 A_{JT9} & = & 0 \ ,
  \nonumber \\
  A_{\cE \cE 2} + A_{\cE T2} & = & 0 \ , \nonumber \\
  3 A_{J\cE 1} + 2 A_{JT1} + 3 A_{JT6} & = & 0 \ , \nonumber \\
  A_{\cE T1} - A_{J\cE 1} & = & 0 \ , \nonumber \\
  3 A_{\cE \cE 1} -2 A_{JT1} - 3 A_{JT6} & = & 0 \ , \nonumber \\
  3 A_{J\cE 2} + A_{JT11} + 2 A_{JT2} + 3 A_{JT7} & = & 0 \ ,
  \nonumber \\
  3 A_{J\cE 3} + A_{JT12} + 2 A_{JT3} + 3 A_{JT8} & = & 0 \ .
  \label{eq:HL-diff-Weyl-relations}
\end{eqnarray}

The numerically-computed values of the coefficients satisfy the
relations (\ref{eq:HL-diff-Weyl-relations}) --- the Ward identities
are preserved at the two-point function level.

\subsection{Scale dependence}

We turn our attention to the derivatives of the two-point functions with
respect to the regularization scale. These are local, so it makes
sense to use the test function procedure.

The derivatives of the correlation functions
\begin{equation}
  \epsilon \frac{\partial}{\partial \epsilon} \langle P^i(x, \tau)
  P^j(0, 0) \rangle = \left[ \delta^{ij} \left( \frac{1}{\epsilon^2}
    A^{\log}_{PP1} + A^{\log}_{PP2} \partial^2 \right) +
    A^{\log}_{PP3} \partial^i \partial^j \right] \delta^{(4)}
  (x, \tau) \ , \label{eq:PPlog}
\end{equation}
\begin{eqnarray}
  \epsilon \frac{\partial}{\partial \epsilon} \langle \cE(x, \tau)
  \cE(0, 0) \rangle & = & \bigg[ \frac{1}{\epsilon^6}
    A^{\log}_{\cE \cE 1} + \frac{1}{\epsilon^4}
    A^{\log}_{\cE \cE 2} \partial^2 + \frac{1}{\epsilon^2} A^{\log}_{\cE
      \cE 3} (\partial^2)^2 + A^{\log}_{\cE \cE 4} (\partial^2)^3 +
    \nonumber \\
    && {} + A^{\log}_{\cE \cE 5} \partial_\tau^2 \bigg] \delta^{(4)}
  (x, \tau) \label{eq:EElog}
\end{eqnarray}
are indeed non-zero and in particular $A^{\log}_{PP3}$ is independent
of $\alpha$, $\beta$ and $\gamma$ (see Appendix
\ref{sec:count-term-coeff}). Therefore, the correlation functions must
have a $\log (\mu \epsilon)$ term regardless of the values of the
couplings, signaling the existence of a type B Weyl anomaly in the
theory.

\section{Discussion} \label{s:discussion}
We have shown through an explicit computation that the $z=3$ Lifshitz scalar has a type B Weyl anomaly when coupled to
a background of Ho{\v r}ava-Lifshitz gravity introduced in~\cite{Horava:2009uw}.  We suspect a similar result will hold in pure HLG itself.  
Experience with relativistic theories would suggest that exact Weyl invariance only comes at the price of perturbative 
unitarity.\footnote{It has been shown that conformal supergravity coupled to a certain super 
Yang-Mills theory has no Weyl anomaly~\cite{Fradkin:1983tg};  however, that theory is haunted by the ghosts of conformal graivty.  
Moreover, it has been argued in~\cite{Cappelli:1990yc} that unitarity implies positivity of the anomaly coefficient $a$ in~(\ref{eq:4anomagain}).}
It is conceivable, though perhaps unlikely, that anisotropic gravity might evade the positivity requirements.  In that case, a modification of
HLG involving  additional degrees of freedom and gauge symmetries (such as might follow from considering some anisotropic locally
supersymmetric theory) could be Weyl invariant. Such developments might be interesting, but the idea is uncomfortably reminiscent of the 
familiar ``Stone Soup'' tale.

The computational technique we use is difficult to extend to
three-point functions, which would be required to find a type A
anomaly.  It would be useful to develop more powerful techniques for
flat space computations.  A more ambitious and difficult undertaking
would be to generalize the point-splitting techniques reviewed
in~\cite{Birrell:1982ix} to the anisotropic case and thereby compute
the renormalized $T_{ij}$, $\cE$ and $P_i$ in an arbitrary HLG
background.  Together with results on chiral anomalies, such as
obtained in~\cite{Dhar:2009dx}, these will yield important constraints
on the structure of correlators of conserved currents in Lifshitz
theories, analogous to those obtained in~\cite{Osborn:1993cr} for
conformal theories.

A natural step in studying the structure of anomalies is to consider the Wess-Zumino consistency conditions.\footnote{These are concisely reviewed in 
the context of Weyl anomalies in~\cite{Bonora:1985cq}.}  We will end our work with a brief look at the consistency condition for Weyl invariance.  
Consider the one-loop effective action $\cW[g,N,\nu]$ and assume there exists a regulator that preserves diffeomorphism and time reparamet\-rization
invariances.  On general grounds the Weyl variation of $\cW$ is given by a local functional
\begin{equation}
\delta_\omega \cW = \int d\tau d^3 x~\sqrt{g}\nu \cA~ \omega,
\end{equation}
where the local function $\cA$ transforms as a scalar under diffeomorphisms and time\-reparametrizations and satisfies the Wess-Zumino
consistency condition: 
\begin{equation}
 [\delta_{\omega_1},\delta_{\omega_2} ] \cW = \int d\tau d^3x~\sqrt{g}\nu \left[ \omega_2 \delta_{\omega_1} - \omega_1 \delta_{\omega_2}\right]\cA=0. 
\end{equation}
In Lorentz-invariant theories this is a constraining requirement.  For example, in $d=4$ it allows just three independent purely gravitational 
parity-invariant terms in $\cA$~\cite{Bonora:1983ff}: 
\begin{equation}
\label{eq:4anomagain}
\cA = -a (\text{Euler}) +  c (\text{Weyl})^2+b \Box R.
\end{equation}

In the anisotropic HLG theory many more terms are possible.  Let
$\Rt_{ij}$ be the Ricci tensor constructed from the
Weyl-invariant metric $\gt = \nu^{-2/3} g$.  Clearly, any scalar $\cA$
of the schematic form
\begin{equation}
\cA = \nablat \Rt \nablat \Rt + \Rt \nablat \nablat \Rt + \Rt \Rt \Rt, 
\end{equation}
where the indices are contracted with the metric $g$, will trivially satisfy the
consistency condition.  These include the square of the Cotton tensor but
clearly contain additional terms.  It would be interesting to
classify these and to determine which, if any, may be eliminated by
local counter-terms.

\acknowledgments We thank A.~Casher, A.~Degeratu, Y.~Oz, A.~Schwimmer
and S.~Yankielowicz for useful discussions. I.A.\ thanks Tel-Aviv
University for hospitality while some of this work was being done.
This work was supported in part by the German-Israeli Project
cooperation (DIP H.52) and the German-Israeli Fund (GIF).

\appendix

\section{The regularized anisotropic
  $\delta$-function} \label{sec:useful-formulas}

We show here that (\ref{eq:HL-Dirac-delta}) satisfies
\begin{equation}
  I = \int d\tau d^3 x \, \delta_\epsilon (\tau, x) = 1 ,
\end{equation}
as required by a representation of Dirac's $\delta$-function. Changing
variables $x = \epsilon u$, $\tau = \epsilon^3 t$ and doing the
angular integration and then changing variables again $v = u^2 (t^2 +
1)^{-1/3}$ to disentangle the $t$ and $v$ variables yields
\begin{equation}
  I = \frac{1}{9 \pi} \int_{-\infty}^\infty \frac{dt}{(t^2 + 1)^{3/2}}
  \int_0^\infty dv \sqrt{v} \left( v \frac{\partial}{\partial v} + 3
  \right) F(v) \ .
\end{equation}
Plugging in the result of the $t$ integration and integrating by parts
in $v$, we have
\begin{equation}
  I = \frac{1}{3 \pi} \int_0^\infty dv v^{1/2} F(v) = 
  \frac{2}{9 \pi} \int_0^\infty dv \, F(v^{2/3})  \ .
\end{equation}
Fortunately, $F(v^{2/3})$ has a nice integral representation
\begin{equation}
  F(v^{2/3}) = \lim_{p \to 1/3} \int_0^\infty dy \frac{\sin \left[ y^{1/3} v^{p}
      \right]}{y^{1/3} v^p} e^{-y} \ .
\end{equation}
Interchanging the $v$ and $y$ integrations and using the integral
\begin{equation}
  \int_0^\infty dx \, \frac{\sin(a x^p)}{a x^p} = \frac{\sqrt{\pi} 2^{(1 -
      2p) / p} a^{-1/p} \Gamma(\frac{1}{2p})}{p \Gamma(\frac{3p -
      1}{2p})} \ , \quad p > 1 \ , a > 0 
\end{equation}
to analytically continue to $p = 1/3$, one finally finds $I = 1$.

\section{Counter-term coefficients} \label{sec:count-term-coeff}
In this section we give the results of numeric computations for the contact term coefficients arising in section~\ref{s:anomaly}. 
In the coefficients that follow we have extracted an over-all factor of $10^{-3}$.

\begin{equation}
A_{\cE\cE 1} = 1.41, \qquad A_{\cE\cE 2} = \ff{1}{4} A_{\cE\cE 1}, \qquad A_{\cE T1 } = -A_{\cE\cE 1}, \qquad A_{\cE T 2} = -\ff{1}{4} A_{\cE\cE 1};
\end{equation}
\begin{equation}
  A_{JP1} = 2.02 \ , \quad A_{JP2} = -0.14 \ , \quad A_{JP3} = 0.63 \ ;
\end{equation}
\begin{eqnarray}
A_{J\cE 1} &=& - A_{\cE\cE 1},\qquad A_{J\cE5} = -\ff{1}{2} A_{\cE\cE 1}, \nonumber\\
A_{J\cE 2} & = & -0.02-0.79\alpha-2.46\beta-1.35\gamma, \nonumber\\
A_{J\cE 3} & = & 0.35 +0.79\alpha-1.40\beta-0.09\gamma, \nonumber\\
A_{J\cE 4} & = & 0.30 +1.24\alpha-0.62\beta+0.26\gamma;
\end{eqnarray}
\begin{eqnarray}
A_{JT1} & = & 2.53,~ \qquad A_{JT 2} = 0.22, \qquad A_{JT3} =-0.04, \nonumber\\
A_{JT4} & = & 0.27,\qquad A_{JT5} = 0.63, \qquad A_{JT6}=-A_{JT10} = -0.28, \nonumber\\
A_{JT7} & = & 0.02-0.70\alpha+1.81\beta+0.14\gamma,\nonumber\\
A_{JT8} & = & -0.35-1.96\alpha+1.33\beta-0.64\gamma, \nonumber\\
A_{JT9} & = & -0.30-1.98\alpha+0.81\beta-0.51\gamma, \nonumber\\
A_{JT11} & = & 4.03\alpha+1.52\beta+3.19\gamma, \nonumber\\
A_{JT12} & = & 3.60\alpha+0.30\beta+2.28\gamma, \nonumber\\
A_{JT13} &=& 1.70\alpha-1.10\beta+0.22\gamma.
\end{eqnarray}

In the logarithmic derivatives we get
\begin{eqnarray}
  A^{\log}_{PP1} & = & -4.04 \ , \nonumber \\
  A^{\log}_{PP2} & = & -3.80 \ , \nonumber \\
  A^{\log}_{PP3} & = & -1.27 \ , \nonumber \\
  A^{\log}_{\cE \cE 1} & = & -6.33 + 4.22 (\alpha + \beta + \gamma) -
    6.33 \alpha^2 -12.67 \alpha \beta - 6.33 \beta^2 - 80.21 \alpha
    \gamma + \nonumber \\
    && {} + 4.22 \beta \gamma - 56.99 \gamma^2 \ ,
  \nonumber \\
  A^{\log}_{\cE \cE 2} & = & -4.47 - 2.83 \alpha + 3.07 \beta - 0.86
    \gamma + 11.32 \alpha^2 + 8.89 \alpha \beta - 2.44 \beta^2 -
    \nonumber \\
    && {} - 78.27 \alpha \gamma + 28.38 \beta \gamma - 65.51 \gamma^2
    \ , \nonumber \\
  A^{\log}_{\cE \cE 3} & = & -2.17 - 7.15 \alpha + 3.62 \beta - 1.77
    \gamma + 2.59 \alpha^2 + 8.33 \alpha \beta - 3.69 \beta^2 -
    \nonumber \\
    && {} - 54.66 \alpha \gamma + 14.58 \beta \gamma - 43.82 \gamma^2
  \ , \nonumber \\
  A^{\log}_{\cE \cE 4} & = & -1.20 - 5.42 \alpha + 2.45 \beta - 0.27
    \gamma - 13.40 \alpha^2 + 2.37 \alpha \beta - 1.13 \beta^2 -
    \nonumber \\
    && {} - 31.02 \alpha \gamma + 3.54 \beta \gamma - 16.81 \gamma^2
  \ , \nonumber \\
  A^{\log}_{\cE \cE 5} & = & 0 \ .
\end{eqnarray}


\begin{thebibliography}{10}

\bibitem{Horava:2009uw}
P.~Ho{\v r}ava, ``{Quantum Gravity at a Lifshitz Point},''
  \href{http://dx.doi.org/10.1103/PhysRevD.79.084008}{{\em Phys. Rev.} {\bf
  D79} (2009)  084008},
\href{http://arxiv.org/abs/0901.3775}{{\tt arXiv:0901.3775 [hep-th]}}.

\bibitem{Deser:1993yx}
S.~Deser and A.~Schwimmer, ``{Geometric classification of conformal anomalies
  in arbitrary dimensions},''
  \href{http://dx.doi.org/10.1016/0370-2693(93)90934-A}{{\em Phys. Lett.} {\bf
  B309} (1993)  279--284},
\href{http://arxiv.org/abs/hep-th/9302047}{{\tt arXiv:hep-th/9302047}}.

\bibitem{Green:1987sp}
M.~Green, J.~Schwarz, and E.~Witten, {\em Superstring Theory, Volume 1}.
\newblock Cambridge University Press, 1987.

\bibitem{Osborn:1993cr}
H.~Osborn and A.~C. Petkou, ``{Implications of Conformal Invariance in Field
  Theories for General Dimensions},''
  \href{http://dx.doi.org/10.1006/aphy.1994.1045}{{\em Ann. Phys.} {\bf 231}
  (1994)  311--362},
\href{http://arxiv.org/abs/hep-th/9307010}{{\tt arXiv:hep-th/9307010}}.

\bibitem{Birrell:1982ix}
N.~D. Birrell and P.~C.~W. Davies, {\em Quantum Fields in Curved Space}.
\newblock Cambridge University Press, 1982.

\bibitem{Duff:1993wm}
M.~J. Duff, ``{Twenty years of the Weyl anomaly},'' {\em Class. Quant. Grav.}
  {\bf 11} (1994)  1387--1404,
\href{http://arxiv.org/abs/hep-th/9308075}{{\tt arXiv:hep-th/9308075}}.

\bibitem{Carvalho:2009nv}
P.~R.~S. Carvalho and M.~M. Leite, ``{Callan-Symanzik-Lifshitz approach to
  generic competing systems},''
\href{http://arxiv.org/abs/0902.1972}{{\tt arXiv:0902.1972 [hep-th]}}.

\bibitem{Dhar:2009dx}
A.~Dhar, G.~Mandal, and S.~R. Wadia, ``{Asymptotically free four-fermi theory
  in 4 dimensions at the z=3 Lifshitz-like fixed point},''
\href{http://arxiv.org/abs/0905.2928}{{\tt arXiv:0905.2928 [hep-th]}}.

\bibitem{Das:2009ba}
S.~R. Das and G.~Murthy, ``{$CP^{N-1}$ Models at a Lifshitz Point},''
\href{http://arxiv.org/abs/0906.3261}{{\tt arXiv:0906.3261 [hep-th]}}.

\bibitem{Iengo:2009ix}
R.~Iengo, J.~G. Russo, and M.~Serone, ``{Renormalization group in Lifshitz-type
  theories},''
\href{http://arxiv.org/abs/0906.3477}{{\tt arXiv:0906.3477 [hep-th]}}.

\bibitem{Freedman:1991tk}
D.~Z. Freedman, K.~Johnson, and J.~I. Latorre, ``{Differential regularization
  and renormalization: A New method of calculation in quantum field theory},''
\href{http://dx.doi.org/10.1016/0550-3213(92)90240-C}{{\em Nucl. Phys.} {\bf
  B371} (1992)  353--414}.

\bibitem{Freedman:1992gr}
D.~Z. Freedman, K.~Johnson, R.~Munoz-Tapia, and X.~Vilasis-Cardona, ``{A Cutoff
  procedure and counterterms for differential renormalization},''
  \href{http://dx.doi.org/10.1016/0550-3213(93)90225-E}{{\em Nucl. Phys.} {\bf
  B395} (1993)  454--496},
\href{http://arxiv.org/abs/hep-th/9206028}{{\tt arXiv:hep-th/9206028}}.

\bibitem{Fradkin:1983tg}
E.~S. Fradkin and A.~A. Tseytlin, ``{Conformal Anomaly in Weyl Theory and
  Anomaly Free Superconformal Theories},''
\href{http://dx.doi.org/10.1016/0370-2693(84)90668-3}{{\em Phys. Lett.} {\bf
  B134} (1984)  187}.

\bibitem{Cappelli:1990yc}
A.~Cappelli, D.~Friedan, and J.~I. Latorre, ``{C theorem and spectral
  representation},''
\href{http://dx.doi.org/10.1016/0550-3213(91)90102-4}{{\em Nucl. Phys.} {\bf
  B352} (1991)  616--670}.

\bibitem{Bonora:1985cq}
L.~Bonora, P.~Pasti, and M.~Bregola, ``{Weyl Cocycles},''
{\em Class. Quant. Grav.} {\bf 3} (1986)  635.

\bibitem{Bonora:1983ff}
L.~Bonora, P.~Cotta-Ramusino, and C.~Reina, ``{Conformal Anomaly and
  Cohomology},''
\href{http://dx.doi.org/10.1016/0370-2693(83)90169-7}{{\em Phys. Lett.} {\bf
  B126} (1983)  305}.

\end{thebibliography}
\providecommand{\href}[2]{#2}\begingroup\raggedright\endgroup

\end{document}